\pageno=1
\headline={\ifnum\pageno>1 \hss \number\pageno\  \hss \else\hfill \fi}
\nopagenumbers
\vglue 12mm

\vskip 100pt
\centerline{\bf \ {
VERTEX OPERATOR EXTENSION OF CASIMIR ${WA}_{_{N}}$  ALGEBRAS
\footnote{$^{^{_{\dagger}}}$}
{\it{This article is dedicated to the $225^{th}$ anniversary of the
Istanbul Technical University.}}
 }}
\vskip 5mm
\centerline{H.\ T.\ \"OZER \footnote{$^{^{_{\dagger \dagger}}}$}
{e-mail\ :\ ozer @ itu.edu.tr }}

\vskip 5mm
\centerline{\it {Physics Department,\ Faculty of Science and Letters,
Istanbul Technical University
,}}
\centerline{\it { 80626,\ Maslak,\ Istanbul,\ Turkey }}
\vskip 15mm
\noindent
We give an extension of Casimir ${WA}_{_{N}}$ algebras including a vertex
operator which depends on non-simple roots of $A_{_{N-1}}$.

\vskip 30mm

\par\vfill\eject


\noindent{\bf{1.\ Introduction}}
\vskip5mm
\noindent
Conformal symmetry plays an important role in string theory $^{1}$ and also
in statistical physics $^{2-3}$.\ Its underlying symmetry algebra is Virasoro
algebra.\ Casimir ${WA}_{_{N}}$ algebras are the higher spin extensions of this
 algebra $^{4-7}$. The idea to extend the  Casimir ${WA}_{_{N}}$ algebras with
the introduction of a vertex operator also seems to be relevant in  two-dimensional
field  theories.\ This construction was presented first by V.A. Fateev
and A.B. Zamolodchikov  for $A_{_{2}}$ $^{8}$  and there are several works $^{9-11}$
dealing with these constructions in the literature. \ The purpose of this
work is to establish a method in this direction.\ Rather than the potential
fields,\ we tried to extend these
studies for primary fields by calculating explicitly all OPEs  between primary
fields and a vertex operator.\ For this,\ we emphasize here that our Casimir
$W_N$-algebra contains only primary fields.\ We explicitly present our results
for OPEs between the primary fields and a vertex operator corresponding  to a
non-simple root of $A_{_{N-1}}$.\ The starting point in this work
based on  a bosonic vertex operator definition in Ref.[12] .\
We then give a general definition of vertex
operator,\ which depends on non-simple roots of $A_{_{N-1}}$.

This paper is organized as follows\ :\ In Sec. 2.\ we recapitulate primary basis
for Casimir $W_{_N}$ algebra by utilizing known Miura transformation with
Feigin-Fuchs type of free massless scalar fields.In Sec.3,\ we constructed
a vertex operator extension of the  Casimir $W_{_{N}}$ algebras by calculating
explicitly all nontrivial OPEs  between primary fields and a vertex operator.
Finally,\ all these used techniques work only for primary fields.\ This
calculation was performed  in $Mathematica^{TM}$ $^{13,14}$.

\vskip 5mm

\noindent{\bf{ 2. \ The Casimir $W_{_{N}}$ Algebras Basis}}
\vskip 5mm
\noindent
In this section we recapitulate primary basis for the Casimir $W_{_N}$
algebras from the Feigin-Fuchs type of free massless scalar field
realization point of view $^{7-10}$.

The  Casimir $ W_N $ algebra is an associative algebra generated by a set of
chiral currents ${U_{_k} (z)}$, of conformal dimension k~$~(k=1 , \cdots , N) $.
The folloving Miura transformation gives the construction  of Casimir $ W_N $
algebra
$$
R_{_N} (z)=
-\sum_{k=0}^N U_{_k} (z) (\alpha_0 \partial)^{N-k}=
: \prod_{j=1}^N (\alpha_{_0} \partial_{_z}- h_{_j}(z) ):,
\eqno(2.1)
$$
where  symbol $ : ~ ~:$  shows normal ordering .\ Here
$\alpha_{_0}$ is a free parameter. \ $ \varphi(z) $ has $ N-1 $ component
which are Feigin Fuchs-type of free massless scalar fields.\ This transformation
determines completely the fields $\{U_{_k} (z)\}$ with
$$
h_{_j}(z)=i  \mu_{_j}\partial \varphi(z)
\eqno(2.2)
$$
Here,\ ${\mu}_i $'s, $(i=1 , \cdots , N) $ are the weights
of the fundamental (vector) representation of  $A_{_{N-1}} $,~satisfying
$ \sum_{i=1}^N {\mu}_i=0 $ and ~${\mu}_i .{\mu}_j=\delta_{_{ij}}-{1\over N} $.
 \ The simple roots of
 $A_{_{N-1}} $ are given by $ {\alpha}_i={\mu}_i-{\mu}_{i+1} $, $(i=1 , \cdots , N-1) $. \ The Weyl vector
of  $A_{_{N-1}} $ is denoted as
$ \rho={{1} \over {2}} \sum_{\alpha > 0} {\bf \alpha^{+}} $ where $\alpha^{+}$ are
the positive roots of  $A_{_{N-1}} $. \ A free scalar field $ \varphi (z) $
is a single-valued function on the complex plane and its mode expansion
is given by
$$
i\,\partial{\varphi}(z)=\sum_{n \in Z} a_n z^{-n-1}.
\eqno(2.3)
$$
Canonical quantization gives the commutator relations
$$
[a_m,a_n]=m \delta_{_{m+n,0}}\ ,
\eqno(2.4)
$$
and these commutator relations are equivalent to the contraction
$$
 \partial{\varphi}(\underline{z)\partial{\varphi}(}w) =-{1 \over {(z-w)^2}} \ .
\eqno(2.5)
$$
By using single contraction $ \partial{\varphi}(\underline{z)\partial
{\varphi}(}w) $,\ a contraction of $ h_{_j}(z)$ with itself is given by
$$
h_{_i}(\underline{z)\,h_{_j}(}w)={{\delta_{_{ij}}-{1\over N}}\over {(z-w)^2}}
\eqno(2.6)
$$
The fields $ \{ U_k (z)\} $ can be obtained by expanding $R_{_N} (z)$.\
We present  a first few one as in the following
$$
U_{_0}(z)=-1~,~~U_{_1}(z)=\sum_{_i}\,h_{_i}(z)=0~,~~
U_{_2}(z)=-\sum_{i<j}\,(h_{_i}\,h_{_j})(z)+\alpha_{_0}\,\sum_{_i}(i-1)\,
\partial{h}_{_i}(z)
\eqno(2.7)
$$
One can see that $ U_{_2} (z) \equiv T(z) $ has spin-2,~ which
is called the stress-energy tensor,~ $ U_k (z) $ has spin-k.
\ The standard OPE of $ T(z) $ with itself is
$$ T(z) T(w)={{c/2} \over (z-w)^4} +
{{2 T(w)} \over (z-w)^2}+{{\partial{T}(w)} \over {z-w}}+\cdots
\eqno(2.8)
$$
where the central charge,~for  $A_{_{N-1}} $,~is given by
$$
c=(N-1)~(1-N(N+1) {\alpha_0}^2).
\eqno(2.9)
$$
A primary field $\phi_{_h}(z)$ with conformal spin-h must provide the following
OPE with $T(z)$
$$
T(z)\phi_{_h} (w)={h\,\phi_{_h} (w) \over {(z-w)}^2}
+{{\partial\,\phi_{_h} (w)} \over {(z-w)}}+\cdots
\eqno(2.10)
$$
Therefore the fields $ \{ U_k (z) \} $ are not primary because
$$
\eqalign{
T(z)U_{_k}(w)&={1 \over 2}\!\sum_{s=1}^k {(N\!-\!k\!+\!s)! \over (N\!-\!k)!}
a_{_0}^{s\!-\!2}\Big(((s\!-\!1)(N\!-\!1)+2(k\!-\!1))a_{_0}^2\!-\!{{s\!-\!1}
\over N}\Big) {{U_{_{k\!-\!s}}(w)} \over (z\!-\!w)^{s\!+\!2}}\cr
&~~~~~~~+{{k U_{_k}(w)} \over (z-w)^2}+{{\partial{U_{_k}(w)}} \over
(z-w)}+(TU_k)(z)+\cdots \cr}
\eqno(2.11)
$$
Using above OPE,\ we defined  $^{7}$ (see also $^{6,8-10}$) some primary fields
of the Casimir $ W_N $ algebra ,whose relations are given by

$$
\overline{U}_{_3}(z) =U_3 (z)-{(N-2) \over 2}~ \alpha_{_0}\partial{T}(z)
\eqno(2.12)
$$

$$
\overline{U}_{_4} (z)=
U_{4} (z)+
\Omega_{_{\partial U_{3} }} \partial U_{3} (z)+
\Omega_{_{\partial^2 T}} \partial^2 T(z)+
\Omega_{_{TT}} (TT)(z)
\eqno(2.13)
$$
where
$$
\Omega_{_{\partial U_{3} }} =-{(N-3) \over 2}\,a_{_0}
$$
$$
\Omega_{_{\partial^2 T}} ={(N-2)(N-3) \over {4N(22+5c)}}\,\Big[-3+N(13+3N+2c)\,a_{_0}^2\Big]
$$
$$
\Omega_{_{TT}}={(N-2)(N-3) \over {2N(22+5c)}}\,\Big[5-N(5N+7)\,a_{_0}^2\Big]
\eqno(2.14)
$$
and
$$
\overline{U}_{_5}(z)=
U_{_5}(z)+
\Omega_{_{\partial{U_{_4}}}}\,\partial{U_{_4}}(z) \ +
\Omega_{_{ \partial^{_2} {U_{_3}}} }\,\partial^{2} {U_{_3}}(z) \ +
\Omega_{_{ \partial^{_3} {U_{_2}}} }\,\partial^{3} {U_{_2}}(z) \ +
$$
$$
\Omega_{_{ U_{_2}U_{_3}}}\,(U_{_2}U_{_3})(z) \ +
\Omega_{_{ U_{_2} \partial{U_{_2}}}}\,(U_{_2} \partial{U_{_2}})(z)
\eqno(2.15)
$$
where
$$
\Omega_{_{\partial{U_{_4}}}}=-{(N-4) \over 2}\,a_{_0}
$$
$$
\Omega_{_{ \partial^{_2} {U_{_3}}} } =
{3 \over 4}\,{ (N-3)(N-4) \over {N(114+7c)}}\,\Big[-2+N(20+c+2N)\,a_{_0}^2\Big]
$$
$$
\Omega_{_{ \partial^{_3} {U_{_2}}} } =
{(N-2)(N-3)(N-4)\,a_{_0} \over {12N(114+7c)}}\,\Big[9-N(33+c+9N)\,a_{_0}^2\Big]
$$
$$
\Omega_{_{ U_{_2}U_{_3}}}=
{(N-3)(N-4) \over {N(114+7c)}}\,\Big[7-N(13+7N)\,a_{_0}^2\Big]
$$
$$
\Omega_{_{ U_{_2} \partial{U_{_2}}}}=
{ (N-2)(N-3)(N-4)\,a_{_0} \over {2N(114+7c)}}\,\Big[-7+N(13+7N) \,a_{_0}^2\Big]
\eqno(2.16)
$$
As being in line with ref.15,\ we have calculated all OPEs of the Casimir
$ W_4 $ algebra $^7$ in a previous work .

\vskip 5mm
\noindent{\bf{ 3.\ Operator Product Expansions\ (OPEs)
for Chiral Vertex Operators}}
\vskip 5mm
\noindent
A chiral vertex operators are defined by

$$
V_{\beta}(z)=:\,e^{i\,\beta\,.\,\varphi(z)}\,:
\eqno(3.1)
$$
Here, a non-simple root $\beta$, for $A_{_{N-1}}$ ,is given by
$$
\beta=\sum_{i=1}^{_{N-1}} m_{_i} \alpha_{_i}
\eqno(3.2)
$$
and  The Fubini-Veneziano field $ \varphi(z)$,~which has conformal spin-0
$$
\varphi(z)=q-i p \ln z + i \sum_{_{n \ne 0}} {1 \over n} a_{_n} z^{-n}
\eqno(3.3)
$$
By using conformal spin-0  contraction $\varphi(\underline{z)\,\varphi(}w)=
-\ln \mid z-w \mid$,\ The standard OPE of $V_{_\beta}(z)$ with  $V_{_{\dot{\beta}}}(w)$

$$
V_{_\beta}(z)\,V_{_{\dot{\beta}}}(w)\,=( z-w)^{\beta\,\dot{\beta}}\,
:V_{_\beta}(z)\,V_{_{\dot{\beta}}}(w):
\eqno(3.4)
$$
We can tell that the operator $V_{_\beta}(z)$ carries a root $\beta$ .From
$$
h_{_j}(z)\,V_{\beta}(w)=
-{{\theta_{_j}}\over {z-w}}  V_{_{\beta}}(w)\,+\,\cdots
\eqno(3.5)
$$
where
$$
\theta_{_j}=
\theta_{_j}(\beta)
\equiv\,
(\beta,\mu_{_{j}})\,
\eqno(3.6)
$$
The OPE with the stress-energy tensor $T(z)$ is

$$
T(z)\,V_{\beta}(w)=
{h(\beta)\over {(z-w)^2}}  V_{_{\beta}}(w) \,+\,
{({\eta_{_1}}^{\beta} V_{_{\beta}})(w)\over {z-w}} \,+\,\cdots
\eqno(3.7)
$$
where~${\eta_{_1}}^{\beta}(z)$
\footnote{*}
{if we take $\beta=\,\alpha_{_i}$,~ a simple root,~ then ~
$({\eta_{_1}}^{\beta} V_{_{\beta}})(z)
=\partial V_{_{\beta}}(z).$},is given by

$$
{\eta_{_1}}^{\beta}(z)=
\sum_{i,j}^N\,
(1-\delta_{_{ij}})\,
\theta_{_i}\,
h_{_j} (z)\,
\eqno(3.8)
$$
Thus the vertex operator  $V_{\beta}(z)$
is a conformal field of spin $h(\beta)$
which is algebraic in $\beta$,is given by
$$
h(\beta)=
-\sum_{i<j}\,\theta_{_i}\theta_{_j}\,+\,
\alpha_{_0}\,\sum_{_i}(i-1)\,\theta_{_i} \equiv U_{2}(\beta)
\eqno(3.9)
$$
For $k>2$,~~similar OPEs between the ${U_{_k} (z)}$ and $V_{\beta}(w)$
could not been calculated explicitly in general root $\beta$ for $A_{_{N-1}}$,
~except the highest order singular term,~since the fields ${U_{_k} (z)}$'s
are not primary,~but primary OPEs will be given later in general form.~
We have
$$
U_{_j}(z)\,V_{\beta}(w)=
{{U_{_j}(\beta)}\over {(z-w)}^j}  V_{_{\beta}}(w)\,+\,less \, sing.\,term's\,\cdots
\eqno(3.10)
$$
Explicitely $U_{_j}(\beta)$'s are some polinomial functions in $\beta$.~
These are found to be
$$
U_{3}(\beta)=
- \sum_{i<j<k}\,\theta_{_i}\theta_{_j}\theta_{_k}\,+\,
2 \alpha_{_0} \sum_{i<j}(i-1)\,\theta_{_i}\theta_{_j}\,
$$
$$
+\, \alpha_{_0} \sum_{i<j}(j-i-1)\,\theta_{_i}\theta_{_j}\,+\,
  \alpha_{_0}^2\,\sum_{_i}(i-1)(i-2)\,\theta_{_i}
\eqno(3.11)
$$

$$
U_{4}(\beta)=
- \sum_{i<j<k<l}\,\theta_{_i}\theta_{_j}\theta_{_k}\theta_{_l}\,+\,
  \alpha_{_0}\, \sum_{i<j<k} (i+j+k-6)\,\theta_{_i}\theta_{_j}\theta_{_k}\,
$$
$$
\,+\,\alpha_{_0}^2\, \sum_{i<j} (-11+6\,i+6\,j-i^2-i\,j -j^2)\,\theta_{_i}\theta_{_j}\,
$$
$$
\,+\,  \alpha_{_0}^3\,\sum_{_i}(i-1)(i-2)(i-3)\,\theta_{_i}
\eqno(3.12)
$$
and
$$
U_{5}(\beta)=
- \sum_{i<j<k<l<m}\,\theta_{_i}\theta_{_j}\theta_{_k}\theta_{_l}\theta_{_m}\,+\,
  \alpha_{_0}\, \sum_{i<j<k<l} (i+j+k+l-10)\,\theta_{_i}\theta_{_j}\theta_{_k}\theta_{l}\,
$$
$$
\,+\,  \alpha_{_0}^2\, \sum_{i<j} (-35+10\,i-i^2+10\,j-i\,j-j^2+10\,k-i\,k-j\,k+k^2)\,\theta_{_i}\theta_{_j}\theta_{k}\,
$$
$$
\,+\,  \alpha_{_0}^3\, \sum_{i<j} (i+j-5)(10-5\,i+i^2-5\,j+j^2)\,\theta_{_i}\theta_{_j}
$$
$$
\,+\,  \alpha_{_0}^4\,\sum_{_i}(i-1)(i-2)(i-3)(i-4)\,\theta_{_i}
\eqno(3.13)
$$
Having found the primary fields $\overline{U}_{_j}(z)$ we are now ready to
compute the primary OPEs.The first result is

$$
\overline{U}_{_3}(z)\,V_{\beta}(w)=
{{\overline{U}_{_3}(\beta)}\over {(z-w)}^3}  V_{_{\beta}}(w)\,+\,
{{(\zeta_{_{\beta}}^{_1} V_{_{\beta}})(w) }\over {(z-w)}^2} \,+\,
{{(\zeta_{_{\beta}}^{_2} V_{_{\beta}})(w) }\over {z-w}} \,+\,
\cdots
\eqno(3.14)
$$
where

$$
\overline{U}_{_3}(\beta)=
- \sum_{i<j<k}\,\theta_{_i}\theta_{_j}\theta_{_k}\,+\,
  \alpha_{_0} \sum_{i<j}(i+j-N-1)\,\theta_{_i}\theta_{_j}\,
$$
$$
\,+\,  \alpha_{_0}^2\,\sum_{_i}(N-i)(i-1)\,\theta_{_i}
\eqno(3.15)
$$

$$
\zeta_{_\beta}^{_1}(z)=
{\alpha_{_0} \over{2}}\sum_{i=1}^{N-1}\,(2\,i-N)\,\gamma_{_i}\sigma_{_j}(z)\,+\,
\sum_{i,j,k=j+1}^{N}\,(1-\delta_{_{ij}}-\delta_{_{ik}})\,\theta_{_j}\theta_{_k} h_{_i}(z)
\eqno(3.16)
$$
and
$$
\zeta_{_\beta}^{_2}(z)=
-\alpha_{_0} \sum_{i=1}^{N-1}\,\tau_{_i}\,\partial \Big(h_{_i}(z)+h_{_{i+1}}(z)\Big)\,-\,
\sum_{i,j,k=j+1}^{N}\,(1-\delta_{_{ij}}-\delta_{_{ik}})\,\theta_{_i}(h_{_j} h_{_k})(z)\,
$$
$$
\,+\,\sum_{i,j,k=j+1}^{N}\,(1-\delta_{_{ij}}-\delta_{_{ik}})\,\theta_{_j}\theta_{_k} \partial h_{_i}(z)
\eqno(3.17)
$$
where
$$
\gamma_{_i}=(\beta,\alpha_{_i})\equiv \theta_{_i}-\theta_{_{i+1}}
\eqno(3.18)
$$
$$
\sigma_{_j}(z)=\sum_{i=1}^{j} h_{_j}(z)
\eqno(3.19)
$$
and
$$
\tau_{_j}=(\beta,\lambda_{_j}) \equiv \sum_{i=1}^{j} \theta_{_i}
\eqno(3.20)
$$
One can repeat all the above computations for $j=4$ and $5$.~
The final results have also been calculated but we will not give the explicit
results here because the expressions are quite long.~For the purpose of
illustration we give only the highest order singular term ;

$$
\overline{U}_{_4} (z)\,V_{\beta}(w)=
{{\overline{U}_{_4}(\beta)}\over {(z-w)}^4}  V_{_{\beta}}(w)\,+\,
less \,sing.\,term's\,\cdots
\eqno(3.21)
$$
where
$$
\overline{U}_{_4}(\beta)=
U_{_4} (\beta)-
3 \,\,\Omega_{_{\partial U_{3} }}  U_{_3} (\beta)+
6 \,\,\Omega_{_{\partial^2 T}} U_{_2}(\beta)+
\Omega_{_{TT}} U_{_2}(\beta)\,\Big(U_{_2}(\beta)+2\Big)
\eqno(3.22)
$$
We also obtained the following composite OPE to use ,
$$
(TT)(z)\,V_{\beta}(w)=
{{h(\beta) \Big(h(\beta)+2\Big)}\over {(z-w)}^4}  V_{_{\beta}}(w)\,+\,
{{\Big(2\,h(\beta)+1\Big)}\over {(z-w)}^3}
({\eta_{_1}}^{\beta}\,V_{_{\beta}})(w)\,
$$
$$
\,\,\,\,\,\,\,+\,
{{2 \,h(\beta)\,(T V_{_{\beta}} )(w)}\over{(z-w)^2}}\,+\,
{{({\eta_{_2}}^{\beta}\, V_{_{\beta}})(w)}\over{(z-w)^2}}\,+\,
{{2 \,h(\beta)\,(\partial T V_{_{\beta}} )(w)}\over{z-w}}\,+\,
{{2\,(T {\eta_{_1}}^{\beta}\, V_{_{\beta}})(w)}\over{z-w}}\,+\,\cdots
\eqno(3.23)
$$
where ${\eta_{_1}}^{\beta}(z)$ is already defined in (3.8) and
${\eta_{_2}}^{\beta}(z)$ \footnote{*}
{if we take $\beta=\alpha_{_i}$,~ a simple root ~
,~then $({\eta_{_2}}^{\beta} V_{_{\beta}})(z)
=\partial^2 V_{_{\beta}}(z).$}
is given by

$$
{\eta_{_2}}^{\beta}(z)=
\Big(
\sum_{i,j}^N\,
(1-\delta_{_{ij}})\,
\theta_{_i}\,
h_{_j} (z)\,
\Big)^2\,
\,+\,
\sum_{i,j}^N\,
(1-\delta_{_{ij}})\,
\theta_{_i}\,
\partial h_{_j} (z)\,
\eqno(3.24)
$$
and finally
$$
\overline{U}_{_5} (z)\,V_{\beta}(w)=
{{\overline{U}_{_5}(\beta)}\over {(z-w)}^5}  V_{_{\beta}}(w)\,+\,
less \,sing.\,term's\,\cdots
\eqno(3.25)
$$
where

$$
\overline{U}_{_5}(\beta)=
U_{_5}(\beta)-
4\,\,\Omega_{_{\partial{U_{_4}}}}\,U_{_4}(\beta) \ +
12\,\,\Omega_{_{ \partial^{_2} {U_{_3}}} }\,U_{_3}(\beta) \ -
24\,\,\Omega_{_{ \partial^{_3} {U_{_2}}} }\,U_{_2}(\beta)
$$
$$
+\,\Omega_{_{ U_{_2}U_{_3}}}\,U_{_3}(\beta)\Big(U_{_2}(\beta)+3\Big) \ -
2\,\,\Omega_{_{ U_{_2} \partial{U_{_2}}}}\,U_{_2}(\beta)\Big(U_{_2}\,(\beta)+3\Big)
\eqno(3.26)
$$
\vskip 5mm

\noindent Here  the highest order singular terms are exposed.

\par To summarize ,\ we have given a systematic algorithm to compute
all nontrivial OPEs between primary fields and a vertex operator
in the  Casimir $W_{_{N}}$ algebras basis.

\vskip 5mm

\par\vfill\eject


\noindent{\bf Acknowledgments}
\vskip 5mm

\noindent The author would like to thank H.\ R.\ Karadayi for his valuable
discussions and excellent guidance throughout this research.\ The calculation
of OPE has been done with the help of Mathematica Package OPEDefs.m $^{13}$.

\vskip 5mm

\noindent{\bf References}

\vskip 5mm

\noindent 1. M.B. Green, J.H. Schwarz and E. Witten, Superstring Theory ,
Vols. 1, 2 (Cambridge Univ. Press. Cambridge. 1987).

\noindent 2. C. Itzykson, H. Saluer and J.B. Zuber, eds, Conformal
Invariance and Applications to Statistical Mechanics
(World Scientific, Singapore, 1988).

\noindent 3. A.A. Belavin, A.M. Polyakov and A.B. Zamolodchikov,
Nucl. Phys. B241 (1984) 333.

\noindent 4. J. Thierry-Mieg, Generalizations of the Sugawara construction,
in : Nonperturbative Quantum Field Theory, eds G.\ 't Hooft et al., Proc.
Cargese School 1987 (Plenum Press, New York, 1988) p.567.

\noindent 5. F. Bais, P. Bouwknegt, M. Surridge and K. Schoutens, Nucl. Phys.
B304 (1988) 348; 371.

\noindent 6. K. Hornfeck, Phys. Lett. B275, 355 (1992).

\noindent 7.\ H.T.\"Ozer,``On the construction of $W_{_N}$-algebras in the form
of $A_{_{N-1}}$ Casimir algebras``.Mod.\ Phys.\ Lett.\ A11,\ 1139 (1996).

\noindent 8.\ V.A. Fateev and A.B. Zamolodchikov, Nucl. Phys. B280 (1987) 644.

\noindent 9. V.A. Fateev and S.L. Lukyanov, Sov. Sci. Rev. A Phys. Vol.15 Part.2 (1990) pp.47-77

\noindent 10. V.A. Fateev and S.L. Lukyanov, Int. J. Mod. Phys. A3 (1988) 507.

\noindent 11. I.K. Kostov,Nucl.Phys.B33 (1988) 559.

\noindent 12.\ B.\ Feigin,\ E.\ Frenkel,\ Integral of motion and quantum
groups,\ hep-th/9310022.

\noindent 13. K.\ Thielemans,"A  ${Mathematica^{TM}}$ package for computing
operator product expansions (OPEdefs 3.1)",\ Theoretical Phys.Group,\ Imperial
College,\ London(UK).

\noindent 14.\ S. Wolfram, ${Mathematica^{TM}}$, (Addison-Wesley,1990).

\noindent 15.\ Blumenhagen et al.,Nucl. Phys. B361. 255 (1991).

\end